\documentclass{elsart}

\usepackage{epsfig}

\usepackage{amssymb}

\begin{document}

\begin{frontmatter}



\title{Nash equilibria in quantum games with generalized two-parameter strategies}


\author{Adrian P. Flitney\corauthref{cor}}
\corauth[cor]{Corresponding author.}
\ead{aflitney@physics.unimelb.edu.au}
\ead[url]{http://aslitney.customer.netspace.net.au/default.html}
\author{Lloyd C. L. Hollenberg}
\address{Centre of Excellence for Quantum Computer Technology and School of Physics,
University of Melbourne, Parkville, VIC 3010, Australia}

\begin{abstract}
In the Eisert protocol for $2 \times 2$ quantum games
[{\em Phys.\ Rev.\ Lett.} {\bf 83}, 3077],
a number of authors have investigated the features arising
from making the strategic space
a two-parameter subset of single qubit unitary operators.
We argue that the new Nash equilibria and the classical-quantum transitions that occur
are simply an artifact of the particular strategy space chosen.
By choosing a different,
but equally plausible,
two-parameter strategic space
we show that different Nash equilibria
with different classical-quantum transitions can arise.
We generalize the two-parameter strategies
and also consider these strategies in a multiplayer setting.
\end{abstract}

\end{frontmatter}

\section{Introduction}
The intersection of the mathematical theory of strategic conflict situations,
known as game theory,
and the tools of quantum mechanics
was first made by Jens Eisert and co-workers~\cite{eisert99}
and by David Meyer~\cite{meyer99}.
Since then a body of theory on quantum games has built up.
The original protocol for two-player, two-strategy ($2 \times 2$) quantum games,
introduced by Eisert {\em et al.}~\cite{eisert99,eisert00},
has remained the main tool
for exploring the properties of quantum games.
In this, the players moves are quantum operators acting on qubits,
with non-classical effects being introduced through entanglement.

In the seminal paper~\cite{eisert99}
and a number of subsequent publications~\cite{iqbal01a,du02a,du02d,chen03a,shimamura04a,ozdemir04a,ozdemir04b}
a particular two-parameter subset of SU(2)
is chosen as the strategic space for the players.
This has the advantage of mathematical simplicity
and, in addition,
new Nash equilibria appear
that are able to solve some of the dilemmas of classical game theory.
For example, in the well known game of Prisoners' Dilemma~\cite{rapoport65}
there is a conflict between the Nash equilibrium (NE),
that determines the strategies that two self-interested rational players would choose to maximize their payoffs,
and the Pareto efficient outcome,
that is the best overall for the players.
With the commonly used two-parameter strategy space
in the Eisert protocol
this dilemma is resolved
with a new NE appearing that coincides with the Pareto optimal (PO) outcome~\cite{eisert99}.
Subsequent investigations revealed three regions
in the space of the entangling parameter:
when the entanglement is greatest there is a quantum region with the new NE,
at minimal entanglements the game behaves classically,
and between these there is a classical-quantum region with intermediate behaviour~\cite{du01b,du03b}.

It was quickly observed by Benjamin and Hayden~\cite{benjamin01a}
that the two-parameter strategies were not the most general,
and suffer from being incomplete under composition.
Three parameters are necessary to describe (up to an arbitrary phase) a general SU(2) operator,
and in the full three-parameter strategy space there is no NE
in deterministic quantum strategies~\cite{benjamin01a}.
Nash equilibria in
mixed quantum strategies exist,
that is, in linear combinations of unitaries~\cite{eisert00,flitney02c}.
However, among these a different dilemma arises:
they form a continuous set and the arrival at a unique NE may be problematic.
In addition, in Prisoners' Dilemma,
the payoff from the mixed quantum NE,
though superior to that from the classical NE,
is still worse than the PO outcome.

This Letter demonstrates that a different,
but equally justifiable,
two-parameter strategic space can give rise to different NE
with different classical-quantum transitions.
This result supports the position that the new NE
and the classical-quantum boundaries to which they give rise
are nothing but artifacts of the particular choice of sub-space.
Indeed, though it is easy to see why allowable choices of the players
may be limited by unitarity,
since this gives the set of reversible quantum operators,
rather than the space of all possible quantum operations,
it is difficult to see a physical justification for the use of two-parameter strategies,
in particular the preference of one two-parameter set over another.

The paper is organized as follows.
Section \ref{s-old2param} summarizes the results for the existing two-parameter strategies
for the game of Prisoners' Dilemma
as well as presenting new results for the games of Chicken and Battle of the Sexes.
In section \ref{s-new2param} we present an alternative two-parameter strategy space
and consider the three games again,
showing quite different behaviour.
In section \ref{s-NPD} we generalize the results for Prisoners' Dilemma to $N$ players
and in section \ref{s-gen2param} we consider two-player games
with generalized two-parameter strategy spaces.

\section{Summary of the Eisert protocol and results for the existing two-parameter strategies}
\label{s-old2param}
There have been many papers summarizing the Eisert protocol for $2 \times 2$ quantum games
so it shall be described here as briefly as possible.
For a more detailed description see, for example, Ref.~\cite{flitney02c}.

The players' choice is encoded in a qubit, initially prepared in the $|0\rangle$ state.
An operator $\hat{J}$,
dependent on a parameter $\gamma \in [0,\pi/2]$,
\begin{equation}
\hat{J} = \cos \frac{\gamma}{2} \hat{I} \otimes \hat{I} \:+\: i \sin \frac{\gamma}{2} \,
		\hat{\sigma}_x \otimes \hat{\sigma}_x,
\end{equation}
entangles the players' qubits to produce the state $(|00\rangle + i |11\rangle)/\sqrt{2}$.
The players' strategies are the choice of local operator that they now make to act on their qubit.
The space of allowable operators forms the strategic space of the players.
Because of the entanglement,
the actions of the players are not independent.
The two classical pure strategies are represented by $\hat{I}$ and $i \hat{\sigma}_x$.
After the players' moves
$\hat{J}^{\dagger}$ is applied to the pair of qubits
with the consequence that if both players choose classical strategies
the result is entirely equivalent to the classical game.
Finally, a measurement is made in the computational basis
and payoffs are awarded using the classical payoff matrix.
The state of the players' qubits prior to the final measurement can be computed by
\begin{equation}
| \psi_f \rangle = \hat{J}^{\dagger} (\hat{A} \otimes \hat{B}) \hat{J} | \psi_i \rangle,
\end{equation}
where $\hat{A}$ and $\hat{B}$ represent Alice's and Bob's moves, respectively.
We are interested in the expectation value $\langle \$ \rangle$ of the players' payoffs.
This can be calculated by
\begin{equation}
\langle \$ \rangle = \$_{00} |\langle \psi_f|00 \rangle|^2 +
			\$_{01} |\langle \psi_f|01 \rangle|^2 +
			\$_{10} |\langle \psi_f|10 \rangle|^2 +
			\$_{11} |\langle \psi_f|11 \rangle|^2,
\end{equation}
where $\$_{ij}$ is the payoff to the player associated with the game outcome
$ij, \; i, j \in \{0,1\}$.

The two parameter quantum strategies of Eisert {\em et al.}~\cite{eisert99}
are drawn from the set\footnote{There are some notational differences with Ref.~\cite{eisert99}
but these are of no significance.}
\begin{eqnarray}
\label{e-old2param}
S^{(1)} &=& \{ \hat{M}^{(1)}(\theta, \phi): \theta \in [0, \pi], \, \phi \in [0, \pi/2] \}, \\ \nonumber
\hat{M}^{(1)}(\theta, \phi) &=& \left( \begin{array}{cc} e^{i \phi} \cos(\theta/2) & i \sin(\theta/2) \\
								  i \sin(\theta/2) & e^{-i \phi} \cos(\theta/2)
					 \end{array} \right).
\end{eqnarray}
A possible parameterization of the full space of SU(2) operators is
\begin{equation}
\label{e-3param}
\hat{M}(\theta, \alpha, \beta) =
	\left( \begin{array}{cc}
		e^{i \alpha} \cos (\theta/2) & i e^{i \beta} \sin (\theta/2) \\
		i e^{-i \beta} \sin (\theta/2) & e^{-i \alpha} \cos(\theta/2)
	    \end{array} \right),
\end{equation} 
where $\theta \in [0, \pi]$ and $\alpha, \beta \in [-\pi, \pi]$.
If we use the latter set for the strategic space
any operation carried out by one player can be reversed by the other.
Consequently any strategy has a counter-strategy~\cite{eisert00,benjamin01a,flitney02c}.
Pure strategy equilibria in the quantum game
will only arise when the classical game,
from which it is derived,
has a purely co-operative solution.
Such games are of little interest to game theory
and, alas, generally do not model any real world situations either!

The two-parameter strategic space of Eq.~(\ref{e-old2param})
can give rise to interesting properties.
As an example consider the famous
Prisoners' Dilemma~\cite{rapoport65}.
Here the players' moves are known as
cooperation ($C$) or defection ($D$).
The payoff matrix can be written as
\begin{equation}
\label{e-PD}
  \begin{array}{c|cc}
	 & \mbox{Bob}: C & \mbox{Bob}: D \\
	\hline
	\mbox{Alice}: C & (3,3) & (0,5) \\
	\mbox{Alice}: D & (5,0) & (1,1)
  \end{array}
\end{equation}
where the numbers in parentheses represent payoffs to Alice
and Bob, respectively.
The game is symmetric and there is a dominant strategy,
that of always defecting,
since it gives a better payoff regardless of the actions of the other player.
The Nash equilibrium is $D \otimes D$
since from this outcome neither player can improve their payoff
by a unilateral change in strategy.
However, $D \otimes D$ is not such a good result for the players
since had they both cooperated
they would have both received a payoff of three rather than one.
The outcome $C \otimes C$ is the PO result,
the one from which neither player can improve their payoff without the other being worse off.
It is this conflict between the NE and PO outcomes that forms the dilemma of the game's name.
In social and political life
this dilemma between individual and group rationality
is responsible for much of the conflict
throughout the world.
In the quantum version of this game
the strategy ``always cooperate'' is represented by $\hat{I}$,
while ``always defect'' is represented by $i \hat{\sigma}_x$.
With maximal entanglement,
$\gamma = \pi/2$,
a new strategy
\begin{equation}
\label{eq-Q}
\hat{C}' = \hat{M}(0, \pi/2) = \left( \begin{array}{cc} i & 0 \\ 0 & -i \end{array} \right)
\end{equation}
emerges as the preferred strategy for both players\footnote{
In Ref.~\cite{eisert99} the strategy $\hat{C}'$ is called $\hat{Q}$.},
with the NE being $\hat{C}' \otimes \hat{C}'$.
The new NE gives an expected payoff of three to both players
there by breaking the dilemma.
The situation for non-maximal entanglement is described by Du {\em et al.}~\cite{du01b}.
Once $\sin^2 \gamma$ slips below $\frac{2}{5}$,
$\hat{C}' \otimes \hat{C}'$ is no longer a NE.
Instead, two new asymmetric NE emerge,
$\hat{C}' \otimes \hat{D}$ and $\hat{D} \otimes \hat{C}'$,
with a payoff of $5 \sin^2 \gamma$ to the player choosing $\hat{C}'$
and $5 \cos^2 \gamma$ to the defecting player.
Finally, for $\sin^2 \gamma < \frac{1}{5}$ the game behaves classically
with a NE of $\hat{D} \otimes \hat{D}$.
These regimes and their corresponding payoffs are indicated in figure~\ref{f-PD}.
In the intermediate region,
the existence of a pair of asymmetric NE each favouring one player over the other
presents a dilemma equivalent to that found in the classical game of Chicken:
How do the players arrive at a unique solution?

Chicken is similar to Prisoners' Dilemma
except that mutual defection is the worst outcome:
\begin{equation}
\label{eq:chicken}
  \begin{array}{c|cc}
	 & \mbox{Bob}: C & \mbox{Bob}: D \\
	\hline
	\mbox{Alice}: C & (3,3) & (1,4) \\
	\mbox{Alice}: D & (4,1) & (0,0)
  \end{array}
\end{equation}
The classical game has two NE in pure strategies,
$C \otimes D$ and $D \otimes C$,
both of which are unsatisfactory for the cooperating player
in comparison with the PO outcome of $C \otimes C$.
A quantum version of this game have been considered in the Eisert protocol~\cite{eisert00}
and some of its features have been examined~\cite{flitney03a}.
With the two parameter strategy set,
the effect on the NE of varying the entanglement parameter
has not previously been published,
though it follows the pattern of Prisoners' Dilemma.
With $\sin^2 \gamma \ge \frac{1}{3}$
the strategy profile $\hat{C}' \otimes \hat{C}'$
is again a NE with payoffs of three to both players.
For $0 < \sin^2 \gamma \le \frac{1}{3}$,
$\hat{C}' \otimes \hat{D}$ and $\hat{D} \otimes \hat{C}'$
are NE with payoffs of $4 - 3 \sin^2 \gamma$ to the $\hat{C}'$ player
and $1 + 3 \sin^2 \gamma$ to the defector.
As $\gamma \rightarrow 0$ these strategy profiles become equivalent to the two classical NE,
as indicated in figure~\ref{f-chicken}.

Another favourite dilemma of game theorists,
and one studied in a number of quantum game papers~\cite{flitney03a,marinatto00,benjamin00,nawaz04a}
is the Battle of the Sexes,
where a couple each have a preferred activity
but are trying to coordinate their actions in the absence of communication.
Alice wants to go to the opera ($O$)
while Bob prefers to watch television ($T$).
The payoff matrix is
\begin{equation}
\label{eq:bos}
  \begin{array}{c|cc}
	 & \mbox{Bob}: O & \mbox{Bob}: T \\
	\hline
	\mbox{Alice}: O & (2,1) & (0,0) \\
	\mbox{Alice}: T & (0,0) & (1,2)
  \end{array}
\end{equation}
The classical game has two NE solutions,
$O \otimes O$ and $T \otimes T$.
The dilemma is similar to that in Chicken
since there is no way to coordinate the decisions of the players and arrive at a unique NE.
If the players choose differently the worst possible result for both players is obtained.
The effect of using the two-parameter strategy set $S^{(1)}$
in this game has not previously been considered.
If Alice plays $\hat{M}^{(1)}(\theta_{\rm A}, \phi_{\rm A})$
while Bob plays $\hat{M}^{(1)}(\theta_{\rm B}, \phi_{\rm B})$,
the probability of the outcome $OO$ is
\begin{equation}
\label{e-pOO}
p_{OO} = [\cos^2 \gamma \, \sin^2 (\phi_{\rm A} + \phi_{\rm B}) \:+\: \cos^2 (\phi_{\rm A} + \phi_{\rm B})] \,
		c^2_{\rm A} c^2_{\rm B},
\end{equation}
while the probability of $TT$ is
\begin{eqnarray}
\label{e-pTT}
p_{TT} &=& \sin^2 \gamma \, \sin^2 (\phi_{\rm A} + \phi_{\rm B}) \,
		c^2_{\rm A} c^2_{\rm B} \:+\: s^2_{\rm A} s^2_{\rm B} \\ \nonumber
 && \makebox[2mm]{} - \frac{1}{2} \sin \gamma \sin \theta _{\rm A}
				\sin \theta_{\rm B} \sin(\phi_{\rm A} + \phi_{\rm B}),
\end{eqnarray}
where $s_k \equiv \sin(\theta_k/2), \; c_k \equiv \cos(\theta_k/2), \; k \in \{{\scriptstyle \rm A, B}\}$.
If $\theta_{\rm A} = 0$, that is, Alice tries to choose her favourite activity of Opera,
then Bob's expected payoff is
\begin{equation}
\langle \$_{\rm B} \rangle =
	\cos^2 \theta_{\rm B} \, [1 \:+\: \sin^2 \gamma \, \sin^2 (\phi_{\rm A} + \phi_{\rm B})].
\end{equation}
Hence, as with the classical scenario,
Bob will want to coordinate his selection of $\theta$ with Alice's
by setting $\theta_{\rm B}=0$.
However, for any $\phi_{\rm A}$ chosen by Alice,
Bob can always select $\phi_{\rm B} = \pi/2 - \phi_{\rm A}$
to give payoffs of
\begin{eqnarray}
\label{e-bos_payoffs}
\langle \$_{\rm A} \rangle = 2 - \sin^2 \gamma; \\ \nonumber
\langle \$_{\rm B} \rangle = 1 + \sin^2 \gamma.
\end{eqnarray}
For $\gamma > \pi/4$ such a result favours Bob over Alice.
Conversely,
if Bob plays $\hat{M}^{(1)}(0, \phi_{\rm B})$
Alice's expected payoff is
\begin{equation}
\langle \$_{\rm A} \rangle =
	\cos^2 \theta_{\rm A} \, [2 \:-\: \sin^2 \gamma \, \sin^2 (\phi_{\rm A} + \phi_{\rm B})].
\end{equation}
Hence Alice will counter with 
$\hat{M}^{(1)}(0, \pi - \phi_{\rm B})$
to ensure a payoff of two compared with one for Bob.
With the restriction in Eq~(\ref{e-old2param}) of $0 \le \phi \le \pi/2$
the best she can be guaranteed of achieving is
$\sin^2 (\phi_{\rm A} + \phi_{\rm B}) \le \frac{1}{2}$,
resulting in
\begin{eqnarray}
\label{e-bos_payoffs1}
\langle \$_{\rm A} \rangle \ge 2 - \frac{1}{2} \sin^2 \gamma; \\ \nonumber
\langle \$_{\rm B} \rangle \le 1 + \frac{1}{2} \sin^2 \gamma.
\end{eqnarray}
There is no equilibrium with $\theta_{\rm A} = \theta_{\rm B} = 0$
since every $\phi$ chosen by one party has an optimal counter by the other.

A cursory examination of Eqs.~(\ref{e-pOO}) and (\ref{e-pTT})
reveal that for $\theta_{\rm B} = \pi$
the phases $\phi_{\rm A}$ and $\phi_{\rm B}$ are not relevant
to the payoffs of Alice or Bob.
The result is a purely classical NE of $\hat{T} \otimes \hat{T}$,
where $\hat{T} = \hat{M}^{(1)}(\pi, \phi)$ for arbitrary $\phi$.
That is, we have started with a symmetric game
and by our choice of strategic space we have a unique NE
with asymmetric payoffs!
This is evidence that the strategy set $S^{(1)}$
has an inherent bias,
an observation that has not been made before.
Further indications of this bias are presented in the next section.

\section{Alternate two-parameter strategy sets}
\label{s-new2param}
To the present authors there seems no reason not to put the two-parameter strategy set
\begin{eqnarray}
\label{e-new2param}
S^{(2)} &=& \{ M(\theta, \phi): \theta \in [0,\pi], \phi \in [0,\pi/2] \}, \\ \nonumber
\hat{M}^{(2)}(\theta, \phi) &=& \left( \begin{array}{cc} \cos(\theta/2) & i e^{i \phi} \sin(\theta/2) \\
								  i e^{-i \phi} \sin(\theta/2) & \cos(\theta/2)
					 \end{array} \right),
\end{eqnarray}
on an equal footing to $S^{(1)}$.
This strategic space has already been used to explore the NE payoff versus entanglement relationship
in two and three player Prisoners' Dilemma~\cite{du02b}
but the authors did not comment on the difference between
their results and earlier ones for two player Prisoners' Dilemma~\cite{du01b,du03b}
found using the strategy space of $S^{(1)}$.
We now explore the consequences of this choice of strategic space on the equilibria
of the Prisoners' Dilemma, Chicken and the Battle of the Sexes.
The results for the Prisoners' Dilemma are the equivalent,
up to a minor notational change in Eq.~(\ref{e-new2param}),
to those given in Ref.~\cite{du02b}.

If Alice plays $\hat{M}^{(2)}(\theta_{\rm A}, \phi_{\rm A})$
and Bob $\hat{M}^{(2)}(\theta_{\rm B}, \phi_{\rm B})$,
the probability of the four possible outcomes of a $2 \times 2$ quantum game
in the Eisert protocol are
\pagebreak
\begin{eqnarray}
p_{CC} &=& \sin^2 \gamma \, \sin^2(\phi_{\rm A} + \phi_{\rm B}) \, s^2_{\rm A} s^2_{\rm B}
		+ c^2_{\rm A} c^2_{\rm B} \\ \nonumber
 && \makebox[2mm]{} +\: \frac{1}{2} \sin \gamma \sin \theta_{\rm A} \sin \theta_{\rm B}
			\sin(\phi_{\rm A} + \phi_{\rm B}), \\ \nonumber
p_{CD} &=& [\cos^2 \gamma \sin^2 \phi_{\rm B} \,+\, \cos^2 \phi_{\rm B}]
		c^2_{\rm A} s^2_{\rm B}
 		\:+\: \sin^2 \gamma \sin^2 \phi_{\rm A} \,
			s^2_{\rm A} c^2_{\rm B} \\ \nonumber
 && \makebox[2mm]{} -\: \frac{1}{2} \sin \gamma \sin \theta_{\rm A} \sin \theta_{\rm B}
			\sin \phi_{\rm A} \cos \phi_{\rm B}, \\ \nonumber
p_{DC} &=& [\cos^2 \gamma \sin^2 \phi_{\rm A} \,+\, \cos^2 \phi_{\rm A}]\,
		s^2_{\rm A} c^2_{\rm B}
		\:+\: \sin^2 \gamma \sin^2 \phi_{\rm B} \,
			c^2_{\rm A} s^2_{\rm B} \\ \nonumber
 && \makebox[2mm]{} -\: \frac{1}{2} \sin \gamma \sin \theta_{\rm A} \sin \theta_{\rm B}
			\cos \phi_{\rm A} \sin \phi_{\rm B}, \\ \nonumber
p_{DD} &=& [\cos^2 \gamma \sin^2 (\phi_{\rm A} + \phi_{\rm B}) \,+\, \cos^2 (\phi_{\rm A} + \phi_{\rm B})] \,
		s^2_{\rm A} s^2_{\rm B}.
\end{eqnarray}
The equivalent to $\hat{C}'$ in the new strategy space is
\begin{equation}
\hat{D}' \equiv \hat{M}^{(2)}(\pi, \pi/4)
	   = \frac{1}{\sqrt{2}} \left( \begin{array}{cc} 0 & i - 1 \\ i + 1 & 0 \end{array} \right).
\end{equation}

Consider the expected payoffs when Bob plays $M^{(1)}(\theta, \phi)$ while Alice plays $\hat{D}'$:
\begin{eqnarray}
\langle \$ \rangle &=& \$_{CC} \, \sin^2 \frac{\theta}{2} \, \sin^2 \gamma \, (1 \,-\, \sin 2 \phi)/2 \\ \nonumber
 && \makebox[2mm]{} + \$_{CD} \, \cos^2 \frac{\theta}{2} \, (\sin^2 \gamma)/2 \\ \nonumber
 && \makebox[2mm]{} + \$_{DC} \, \cos^2 \frac{\theta}{2} \left[1 \:-\: \frac{1}{2} \sin^2 \gamma \right] \\ \nonumber
 && \makebox[2mm]{} + \$_{DD} \, \sin^2 \frac{\theta}{2}
	\left[1 \:-\: \frac{1}{2} \sin^2 \gamma \, (1 \,+\, \sin 2 \phi) \right].
\end{eqnarray}
For the Prisoners' Dilemma payoffs of Eq.~({\ref{e-PD}),
Bob maximizes his results for all $\gamma$
by setting $\theta = \phi$ and $\phi = \pi/4$.
Thus $\hat{D}' \otimes \hat{D}'$ is a symmetric NE for all $\gamma$ with
\begin{equation}
\langle \$_{\rm A} \rangle = \langle \$_{\rm B} \rangle = 1 + 2 \sin^2 \gamma.
\end{equation}
The payoff is superior to that of the classical NE of $\hat{D} \otimes \hat{D}$
provided $\gamma > 0$,
as indicated in figure~\ref{f-PD}.
The NE is not strict since there is some flexibility in the allowable values of $\phi$.
The permitted values of $\phi_{\rm A}, \phi_{\rm B}$
are those for which
$\phi_{\rm A} + \phi_{\rm B} = \pi/2$
and both $\phi$ satisfy
\begin{equation}
\frac{3 \sin^2 \gamma \:-\: 1}{5 \sin^2 \gamma}
	< \sin^2 \phi
	< \frac{2 \sin^2 \gamma \:+\: 1}{5 \sin^2 \gamma}.
\end{equation}
However, $\phi_{\rm A} = \phi_{\rm B} = \pi/4$ is a focal point~\cite{schelling60}
that will attract the players for psychological reasons.
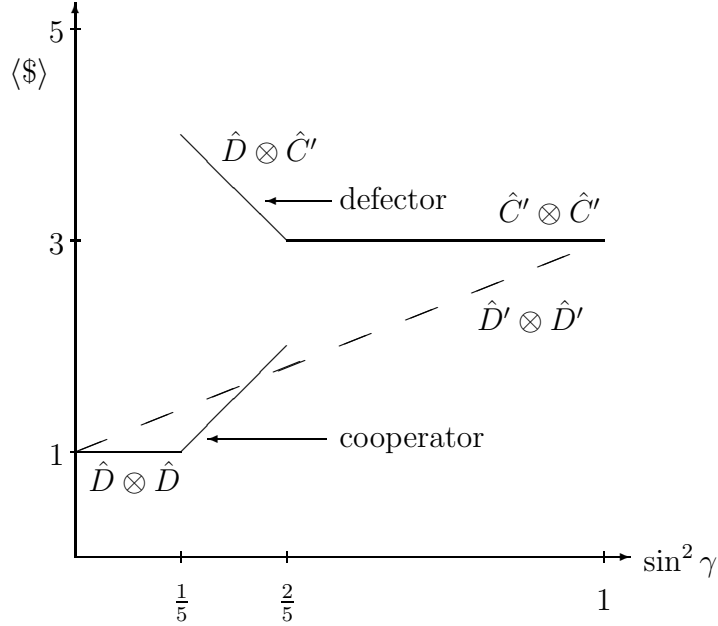
\begin{figure}
\begin{center}
\begin{picture}(230,250)(0,0)
	\put(20,20){\vector(1,0){210}}
	\put(20,20){\vector(0,1){210}}
	\put(235,15){$\sin^2 \gamma$}
	\put(-5,200){$\langle \$ \rangle$}
	\put(60,18){\line(0,1){4}}
	\put(57,0){$\frac{1}{5}$}
	\put(100,18){\line(0,1){4}}
	\put(97,0){$\frac{2}{5}$}
	\put(220,18){\line(0,1){4}}
	\put(217,0){1}
	\put(18,60){\line(1,0){4}}
	\put(10,55){1}
	\put(18,140){\line(1,0){4}}
	\put(10,135){3}
	\put(18,220){\line(1,0){4}}
	\put(10,215){5}

	\put(20,60){\line(1,0){40}}
	\put(60,60){\line(1,1){40}}
	\put(60,180){\line(1,-1){40}}
	\put(100,140){\line(1,0){120}}
	\multiput(20,60)(25,10){8}{\line(5,2){12}}

	\put(25,45){$\hat{D} \otimes \hat{D}$}
	\put(75,170){$\hat{D} \otimes \hat{C}'$}
	\put(180,147){$\hat{C}' \otimes \hat{C}'$}
	\put(172,106){$\hat{D}' \otimes \hat{D}'$}
	\put(117,155){\vector(-1,0){25}}
	\put(120,152){defector}
	\put(115,65){\vector(-1,0){45}}
	\put(120,62){cooperator}
\end{picture}
\caption{The Nash equilibrium payoffs in a quantum version of Prisoners' Dilemma
as a function of the degree of entanglement,
as measured by $\sin^2 \gamma$,
for the strategy space $S^{(1)}$ (solid lines),
and for $S^{(2)}$ (dashed line).
For $S^{(1)}$ there are three regions with different the Nash equilibria:
$\hat{D} \otimes \hat{D}$, $\hat{C}' \otimes \hat{D}$ and $\hat{D} \otimes \hat{C}'$,
and $\hat{C}' \otimes \hat{C}'$.
The strategy $\hat{C}'$ is a cooperative strategy,
becoming equivalent to $\hat{C}$ as $\gamma \rightarrow 0$.}
\label{f-PD}
\end{center}
\end{figure}

In the game of Chicken the strategy profile
$\hat{D}' \otimes \hat{D}'$ results in mutual payoffs of $3 \sin^2 \gamma$.
This is a NE provided $\sin^2 \gamma \ge \frac{1}{3}$.
Again there is some ambiguity in the values of $\phi$
that can give rise to a NE,
but this is unimportant for our purposes\footnote
{For completeness,
we require $\phi_{\rm A} +\phi_{\rm B} = \pi/2$ with both $\phi$ constrained by \\
$\frac{1}{3 \sin^2 \gamma \:+\: 1} < \sin^2 \phi < \frac{3 \sin^2 \gamma}{3 \sin^2 \gamma \:+\: 1}$.}.
Below this level,
the classical strategy profiles
$\hat{C} \otimes \hat{D}$ and $\hat{D} \otimes \hat{C}$
are NE, as indicated in figure~\ref{f-chicken}.
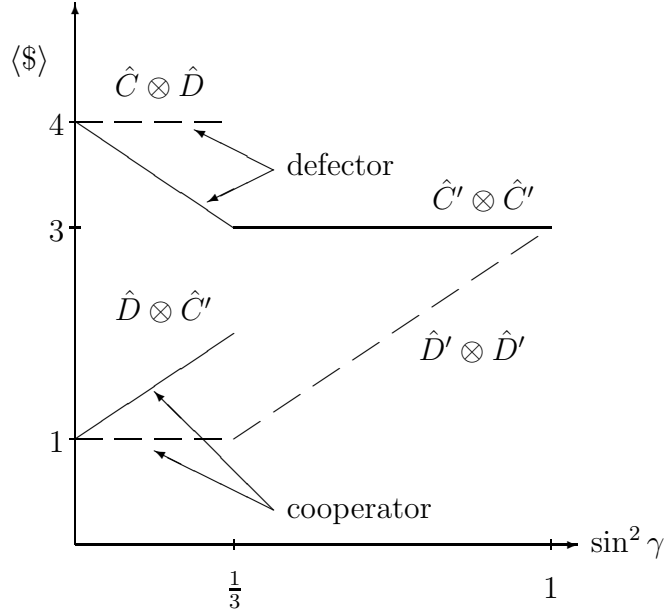
\begin{figure}
\begin{center}
\begin{picture}(205,230)(0,0)
	\put(20,20){\vector(1,0){190}}
	\put(20,20){\vector(0,1){205}}
	\put(215,17){$\sin^2 \gamma$}
	\put(-5,200){$\langle \$ \rangle$}
	\put(80,18){\line(0,1){4}}
	\put(77,0){$\frac{1}{3}$}
	\put(200,18){\line(0,1){4}}
	\put(197,0){1}
	\put(18,60){\line(1,0){4}}
	\put(10,55){1}
	\put(18,140){\line(1,0){4}}
	\put(10,135){3}
	\put(18,180){\line(1,0){4}}
	\put(10,175){4}

	\put(20,60){\line(3,2){60}}
	\put(20,180){\line(3,-2){60}}
	\put(80,140){\line(1,0){120}}
	\multiput(80,60)(15,10){8}{\line(3,2){10}}
	\multiput(20,180)(15,0){4}{\line(1,0){10}}
	\multiput(20,60)(15,0){4}{\line(1,0){10}}

	\put(35,105){$\hat{D} \otimes \hat{C}'$}
	\put(35,190){$\hat{C} \otimes \hat{D}$}
	\put(155,147){$\hat{C}' \otimes \hat{C}'$}
	\put(150,90){$\hat{D}' \otimes \hat{D}'$}
	\put(100,160){defector}
	\put(95,162){\vector(-2,1){30}}
	\put(95,162){\vector(-2,-1){25}}
	\put(100,30){cooperator}
	\put(95,33){\vector(-1,1){45}}
	\put(95,33){\vector(-2,1){45}}
\end{picture}
\caption{The Nash equilibrium payoffs in a quantum version of the game of Chicken
as a function of the degree of entanglement,
as measured by $\sin^2 \gamma$,
for the strategy space $S^{(1)}$ (solid lines),
and for $S^{(2)}$ (dashed line).
In the region $\sin^2 \gamma < \frac{1}{3}$,
the Nash equilibria in the first case are $\hat{C}' \otimes \hat{D}$ and $\hat{D} \otimes \hat{C}'$,
while for $S^{(2)}$ the Nash equilibria are the classical ones
of $\hat{C} \otimes \hat{D}$ and $\hat{D} \otimes \hat{C}$.
The strategy $\hat{C}'$ is a cooperative strategy,
becoming equivalent to $\hat{C}$ as $\gamma \rightarrow 0$.}
\label{f-chicken}
\end{center}
\end{figure}

The reason for the behaviour of the new quantum equilibrium in both
Prisoners' Dilemma and Chicken results from the fact that
$\hat{D}'$ becomes equivalent to $\hat{D}$ as $\gamma \rightarrow 0$.
In the former case this means that $\hat{D}' \otimes \hat{D}'$ remains a NE
for all $\gamma$ since it smoothly asymptotes to the classical NE
of $\hat{D} \otimes \hat{D}$ as $\gamma \rightarrow 0$,
while for Chicken there is a switch to one of the classical NE
$\hat{C} \otimes \hat{D}$ or $\hat{D} \otimes \hat{C}$ for small enough $\gamma$,
since $\hat{D} \otimes \hat{D}$ is an undesirable result for both players.
By comparison,
in the strategic space $S^{(1)}$,
$\hat{C}'$ becomes equivalent to $\hat{C}$ as $\gamma \rightarrow 0$
giving rise to different classical-quantum thresholds
since at some point defection will be favoured by one or both players.

In the Battle of the Sexes,
using the strategic space $S^{(2)}$
reverses the results obtained
for this game presented in the previous section.
Now $\hat{O} \otimes \hat{O}$ is the unique NE,
where $\hat{O} = \hat{M}^{(2)}(0, \phi)$ for arbitrary $\phi$.
The payoffs are
\begin{eqnarray}
\label{e-bos_payoffs2}
\langle \$_{\rm A} \rangle = 1 + \sin^2 \gamma; \\ \nonumber
\langle \$_{\rm B} \rangle = 2 - \sin^2 \gamma,
\end{eqnarray}
the reverse of those in Eq.~(\ref{e-bos_payoffs}).
In the case of both strategic spaces,
the addition of phase factors to only one diagonal
of the matrix for the player operators $M^{(1)}$ or $M^{(2)}$
favours one of the players
by giving them the means to respond to their opponent's preferred choice
while there is no such response to their own desired selection.
Consequently, one of the two classical NE is eliminated.


\section{Nash equilibria in $N$-player quantum Prisoners' Dilemma}
\label{s-NPD}
There is no standard accepted payoff matrix for $N$-player Prisoners' Dilemma,
and for $N > 3$ it is little studied.
However, we consider it of interest in the quantum case since it gives
us some information about $N$-partite entanglement.
Here an $N$-player mutual Prisoners' Dilemma is considered
rather than the more frequently studied case
of a series of two player interactions among multiple players~(e.g., see Ref.~\cite{axelrod84}).
The requirements on the payoffs are that
\begin{itemize}
\item Defection is always the dominant strategy,
that is, a player always receives more for defecting than for cooperating
regardless of the actions of the other players.
\item A player is better off when more of the other players cooperate.
\item When $N-2$ players' moves are fixed,
the remaining $2 \times 2$ game is a two-player Prisoners' Dilemma.
\end{itemize}
These constraints are satisfied if cooperators and defectors receive, respectively,
\begin{eqnarray}
\label{eq-NPD_payoffs}
\$_{C} &=& \left\{ \begin{array}{ll} 0,              & m = 1 \\
						 3 \,+\, 4(m-2), & m > 1, \end{array} \right. \\ \nonumber
\$_{D} &=& 5 \,+\, 4(m-1)
\end{eqnarray}
where $m$ is the number of players cooperating.

With this payoff structure and the strategy space $S^{(1)}$
we find the generalization of the $\hat{C}' \otimes \hat{C}'$ equilibrium in the fully entangled game is
$\hat{C}_{N}^{\,\otimes N}$,
where we have introduced the notation
\begin{equation}
\hat{C}_{n} \equiv \hat{M}^{(1)}(0, \pi/n) = \left( \begin{array}{cc} e^{i \pi/n} & 0 \\ 0 & e^{-i \pi/n} \end{array} \right).
\end{equation}
The strategy $\hat{C}_{N}$ is cooperation ($\theta = 0$)
but with the addition of a phase factor $\pi/N$.
For the strategy set $S^{(1)}$,
all $\hat{D}_{n} \equiv \hat{M}^{(1)}(\pi, \pi/n)$ are equivalent to $\hat{D}$.
To demonstrate that $\hat{C}_{N}$ is a symmetric NE strategy,
consider the payoff to the last player, Larry,
when he plays $\hat{M}^{(1)}(\theta, \phi)$ while all the other players continue with $\hat{C}_{N}$:
\begin{eqnarray}
\langle \$_{\rm L} \rangle &=& \$_{C \ldots C} \,\cos^2 \frac{\theta}{2}
	\, \left[1 \,-\, \frac{1}{2} \sin^2 \gamma \, (1 - \cos(2 \phi - \frac{2 \pi}{N})) \right] \\ \nonumber
 && \makebox[2mm]{} + \$_{C \ldots CD} \,\sin^2 \frac{\theta}{2}
	\, \left[1 \,-\, \frac{1}{2} \sin^2 \gamma \, (1 - \cos \frac{2 \pi}{N}) \right] \\ \nonumber
 && \makebox[2mm]{} + \$_{D \ldots DC} \, \sin^2 \frac{\theta}{2} \,
	\sin^2 \gamma \, (1 - \cos \frac{2 \pi}{N})/2 \\ \nonumber
 && \makebox[2mm]{} + \$_{D \ldots D} \, \cos^2 \frac{\theta}{2} \,
	\sin^2 \gamma \, (1 - \cos(2 \phi - \frac{2 \pi}{N}))/2.
\end{eqnarray}
Since $\$_{C \ldots C} > \$_{D \ldots D}$,
Larry chooses $\phi = \pi/N$ to maximize $\cos(2 \phi - \frac{2 \pi}{N})$.
Given that the result $D \ldots DC$ yields Larry nothing,
\begin{equation}
\langle \$_{\rm L} \rangle = \$_{C \ldots C} \, \cos^2 \frac{\theta}{2}
				\:+\: \$_{C \ldots CD} \, \sin^2 \frac{\theta}{2}
				\, \left[1 \,-\, \frac{1}{2} \sin^2 \gamma \, (1 - \cos \frac{2 \pi}{N}) \right].
\end{equation}
Hence, Larry prefers $\theta = 0$ provided
\begin{equation}
\$_{C \ldots C} \ge \$_{C \ldots CD} \, [1 \,-\, \frac{1}{2} \sin^2 \gamma \, (1 - \cos \frac{2 \pi}{N})].
\end{equation}
Using Eq.~(\ref{eq-NPD_payoffs}) this reduces to
\begin{equation}
\label{eq-threshold}
\sin^2 \gamma \ge \frac{4}{(4 N - 3)(1 - \cos \frac{2 \pi}{N})}.
\end{equation}
Thus for sufficient high entanglement,
$\hat{M}^{(1)}(0, \pi/N) \equiv \hat{C}_{N}$
is the best strategy for Larry
when the strategies of the other players are fixed.
By symmetry this demonstrates that $\hat{C}_{N}^{\, \otimes N}$ is a NE.
When the entanglement parameter $\gamma$ drops below the level given by Eq.~(\ref{eq-threshold})
this strategy profile is no longer a NE.
There is a new NE where one player selects the strategy $\hat{C}_{2}$ while all the others defect.
Again we demonstrate that this is a NE for a certain range of entanglement by considering the payoff
to a player that changes from their equilibrium strategy.
The payoff to Larry when he plays $\hat{M}^{(1)}(\theta, \phi)$ while Alice plays $\hat{C}_{2}$
and the remaining players defect is
\begin{eqnarray}
\langle \$_{\rm L} \rangle &=& \$_{CD \ldots DC} \cos^2 \frac{\theta}{2}
	\, \left[1 \,-\, \frac{1}{2} \sin^2 \gamma \, (1 + \cos(2 \phi) \right] \\ \nonumber
 && \makebox[2mm]{} + \$_{CD \ldots D} \, \sin^2 \frac{\theta}{2} \, \cos^2 \gamma \\ \nonumber
 && \makebox[2mm]{} + \$_{DC \ldots C} \, \sin^2 \frac{\theta}{2} \, \sin^2 \gamma \\ \nonumber
 && \makebox[2mm]{} + \$_{DC \ldots CD} \, \cos^2 \frac{\theta}{2} \,
	\sin^2 \gamma \, (1 + \cos(2 \phi))/2.
\end{eqnarray}
For Larry, $\$_{DC \ldots CD} > \$_{CD \ldots DC}$
so he will choose $\phi = 0$.
Then $\theta = \pi$ is his preferred choice provided
\begin{eqnarray}
\sin^2 \gamma &\le& \frac{ \$_{CD \ldots D} - \$_{CD \ldots DC}}
		       { \$_{DC \ldots CD} + \$_{CD \ldots D} - \$_{CD \ldots DC} - \$_{DC \ldots C} } \\ \nonumber
	      &=& \frac{1}{2} \qquad \forall N,
\end{eqnarray}
where the last line is calculated using the payoffs in Eq.~(\ref{eq-NPD_payoffs}).
So for $\gamma \le \pi/4$ Larry does best by sticking to $\hat{D}$.  
Similarly, if Alice plays $\hat{M}^{(1)}(\theta, \phi)$ while all the other players defect,
her expected payoff is
\begin{eqnarray}
\langle \$_{\rm A} \rangle &=& \$_{CD \ldots D} \, [1 \,-\, \frac{1}{2} \cos^2 \frac{\theta}{2} \,
	\sin^2 \gamma \,(1 - \cos(2 \phi)] \\ \nonumber
 && \makebox[2mm]{} + \$_{DC \ldots C} \, \cos^2 \frac{\theta}{2} \, \sin^2 \gamma \, (1 - \cos 2 \phi)/2 \\ \nonumber
 && \makebox[2mm]{} + \$_{D \ldots D} \, \sin^2 \frac{\theta}{2}.
\end{eqnarray}
Since, for Alice, $\$_{DC \ldots C} > \$_{CD \ldots D}$,
$\phi = \pi/2$ is Alice's preferred value in the range $[0, \pi/2]$.
Then she prefers $\theta = 0$ provided
\begin{eqnarray}
\sin^2 \gamma &\ge& \frac{\$_{D \ldots D} - \$_{CD \ldots D}}{\$_{DC \ldots C} - \$_{CD \ldots D}} \\ \nonumber
              &=& \frac{1}{4 N - 3}.
\end{eqnarray}
Hence for $1/(4 N -3) \le \sin^2 \gamma \le 1/2$
the strategy profile $\hat{C}_2 \otimes \hat{D}^{\otimes N-1}$ is a NE.
There are $N$ equilibria of this form
depending on who plays $\hat{C}_2$.
Of course, the existence of an equilibrium that ``rational'' players should select
avoids the question of how to arrive at such an equilibrium.
The asymmetric nature of this equilibrium would make reaching it extremely problematic in practice.
Where both the asymmetric equilibria and the mutual cooperation equilibrium coexist,
that is, for
\begin{equation}
\frac{4}{(4 N - 3)(1 - \cos \frac{2 \pi}{N})} \le \sin^2 \gamma \le \frac{1}{2},
\end{equation}
mutual cooperation is better for all the players and,
being symmetric,
there is no difficulty in reaching it.
For $\gamma < 1/(4 N -3)$ the only NE is the classical one of mutual defection.
Figure~\ref{f-4PD} shows the NE payoffs for a four-player quantum Prisoners' Dilemma.
The same structure applies for arbitrary $N$,
the main difference being that the line representing the cooperator's payoff in the intermediate region
becomes steeper for increasing $N$.

Straight forward calculations similar to the above demonstrate that there are no
other NEs of the form $\hat{C}_{m}^{\otimes m} \otimes \hat{D}^{\otimes N-m}, \; m=0,1,2,\ldots$,
though we have not ruled out the existence of more complicated equilibria.

When the strategic space is chosen to be $S^{(2)}$ there is a symmetric NE
analogous to that present in the two-player game, $\hat{D}_{2N}^{\otimes N}$,
where
\begin{equation}
\hat{D}_{n} \equiv \hat{M}^{(2)}(\pi,\pi/n) = \left( \begin{array}{cc} 0 & i e^{i \pi/n} \\ i e^{-i \pi/n} & 0 \end{array} \right).
\end{equation}
This is demonstrated to be a NE by showing that any unilateral variation
in strategy produces an inferior outcome for the varying player.
Consider Larry's payoff when he plays $\hat{M}^{(2)}(\theta, \phi)$ while the others continue with $\hat{D}_{2N}$:
\begin{eqnarray}
\langle \$_{\rm L} \rangle &=& \$_{C \ldots C} \, \sin^2 \frac{\theta}{2} \,
	\sin^2 \gamma \, \left[1 \,+\, \cos(2 \phi - \frac{\pi}{N}) \right]/2 \\ \nonumber
 && \makebox[2mm]{} + \$_{C \ldots CD} \, \cos^2 \frac{\theta}{2} \, \sin^2 \gamma \,
 	(1 + \cos \frac{\pi}{N})/2 \\ \nonumber
 && \makebox[2mm]{} + \$_{D \ldots DC} \, \cos^2 \frac{\theta}{2}
 	\, \left[1 \,-\, \frac{1}{2} \sin^2 \gamma \, (1 + \cos \frac{\pi}{N}) \right] \\ \nonumber
 && \makebox[2mm]{} + \$_{D \ldots D} \, \sin^2 \frac{\theta}{2}
	\, \left[1 \,-\, \frac{1}{2} \sin^2 \gamma \, (1 + \cos(2 \phi - \frac{\pi}{N})) \right].
\end{eqnarray}
Since $\$_{C \ldots C} > \$_{D \ldots D}$,
Larry prefers $\phi = \pi/(2 N)$.
Larry will then choose $\theta=\pi$ provided
\begin{equation}
\$_{C \ldots C} \, \sin^2 \gamma \:+\: \$_{D \ldots D}
	\ge \$_{C \ldots CD} \,(1 \,+\, \cos \frac{\pi}{N}).
\end{equation}
With the payoffs given by Eq.~(\ref{eq-NPD_payoffs}),
Larry will select $\theta = \pi$ for all $\gamma$,
provided $N \le 8$.
For $ N \ge 9$ this NE begins to break down at large entanglements,
however, mutual Prisoners' Dilemmas with such large numbers of players
are of little or no practical interest.

Although the payoff structure versus entanglement
presents an interesting picture with its different regions and bifurcations
we would like to emphasize again that the existence of these NE and the associated entanglement thresholds
are a function of the particular strategic space to which the players are restricted.

\begin{figure}
\begin{center}
\epsfig{file=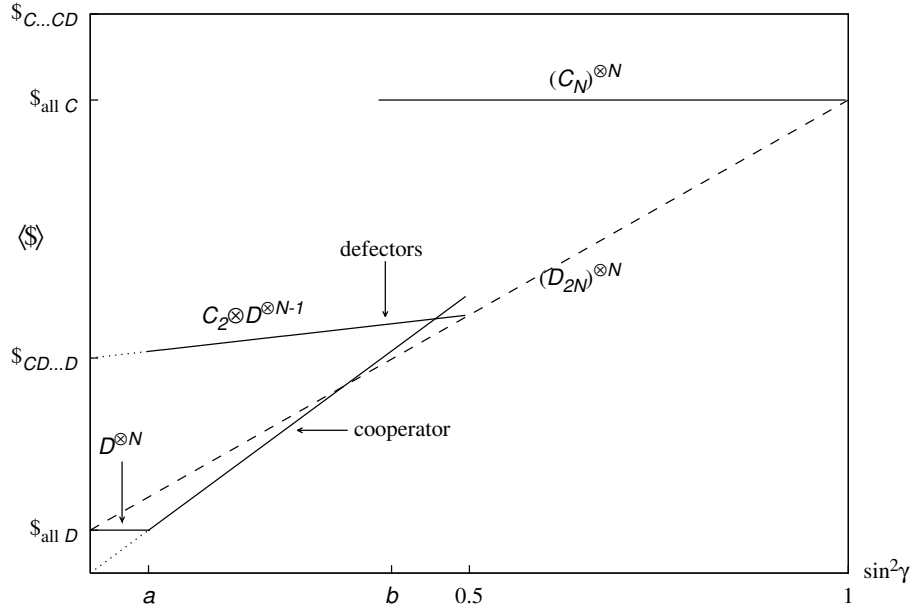, width=12cm}
\end{center}
\caption{The Nash equilibrium payoffs in a quantum version of a four-player Prisoners' Dilemma
as a function of the degree of entanglement,
as measured by $\sin^2 \gamma$,
for the strategic space $S^{(1)}$ (solid lines),
and $S^{(2)}$ (dashed line).
In the case of $S^{(1)}$ there are three regions with different Nash equilibria.
The same structure holds for arbitrary $N$,
the main difference being that the gradient of the line for the cooperator's payoff
in the intermediate region increases with $N$.
The thresholds are, in general, $a = 1/(4 N - 3)$ and $b = 4/((4 N - 3)(1 - \cos \frac{2 \pi}{N}))$.}
\label{f-4PD}
\end{figure}

\section{Two-player quantum games with generalized two-parameter strategies}
\label{s-gen2param}
The strategy spaces $S^{(1)}$ and $S^{(2)}$ can be generalized to
\begin{eqnarray}
\label{e-gen2param}
S^{(1)}_{k} &=& \left( \begin{array}{cc} e^{i \phi} \cos \frac{\theta}{2} & i e^{i k \phi} \sin \frac{\theta}{2} \\
						     i e^{-i k \phi} \sin \frac{\theta}{2} & e^{-i \phi} \cos \frac{\theta}{2}
			     \end{array} \right), \\ \nonumber
S^{(2)}_{k} &=& \left( \begin{array}{cc} e^{i k \phi} \cos \frac{\theta}{2} & i e^{i \phi} \sin \frac{\theta}{2} \\
						     i e^{-i \phi} \sin \frac{\theta}{2} & e^{-i k \phi} \cos \frac{\theta}{2}
			     \end{array} \right),
\end{eqnarray}
where $k \in [0,1]$ is a fixed parameter.
The two strategic spaces are distinct except when $k=1$.
Setting $k=0$ reduces $S^{(j)}_{k} \rightarrow S^{(j)}$ for $j=1,2$.

Now consider a two-player quantum Prisoners' Dilemma using each of the two strategic spaces in Eq.~(\ref{e-gen2param}).
The strategic space $S^{(2)}_{k}$ is uninteresting
since it yields the same payoff versus entanglement structure
as $S^{(2)}$: $\hat{C}' \otimes \hat{C}'$ is not a NE for any $\gamma, k$
while $\hat{D}' \otimes \hat{D}'$ is a NE for all $\gamma, k$
with a payoff to each player of $1 + 2 \sin^2 \gamma$.

For $S^{(1)}_{k}$,
$\hat{C}_{2} \equiv \hat{C}'$ is a symmetric NE strategy
with a payoff of $\$_{CC}$ for a range of $\gamma$ and $k$.
Consider Bob's payoff when he counters $\hat{C}_{2}$
with $\hat{M}^{(1)}_{k}(\pi, \phi)$ for some $\phi$ to be determined:
\begin{equation}
\langle \$_{\rm B} \rangle = \$_{CD} \, (1 - \sin^2 \gamma \, \cos^2 k \phi)
	\:+\: \$_{DC} \, \sin^2 \gamma \, \cos^2 k \phi.
\end{equation}
Bob maximizes this by selecting $\phi = \pi/2$.
He prefers the resulting payoff over $\$_{CC}$ when
\begin{equation}
\label{e-gammak}
\sin^2 \gamma \, \cos \frac{k \pi}{2} < \frac{\$_{CD} - \$_{CC}}{\$_{CD} - \$_{DC}}.
\end{equation}
All $\$_{ij}$ are considered from Bob's perspective,
so using the payoffs of Eq.~(\ref{e-PD}),
the right hand side of the above equation evaluates to $\frac{2}{5}$.
The region satisfying Eq.~(\ref{e-gammak}) is shown in figure~\ref{f-PDk}.

The strategy $\hat{D}_{4k} \equiv \hat{M}^{(1)}_{k}(\pi, \pi/(4k))$, when played by both players,
yields a payoff
\begin{equation}
\label{e-NEk}
\langle \$ \rangle = \$_{CC} \sin^2 \gamma \:+\: \$_{DD} \cos^2 \gamma,
\end{equation}
that, for $\gamma = \pi/2$, is the same as the cooperative payoff.
If we maintain the restriction that $\phi \le \pi/2$ this strategy
can only be chosen when $k \ge \frac{1}{2}$.
The region for which this is a NE can be deduced by considering the payoff to Bob if he instead switches
to $\hat{M}^{(1)}_{k}(0, \phi)$:
\begin{equation}
\langle \$_{\rm B} \rangle = \$_{CD} \, (1 - \sin 2 k \phi)/2
	\:+\: \$_{DC} \left[ 1 \,-\, \sin^2 \gamma \, (1 - \sin 2 k \phi)/2 \right].
\end{equation}
For $k \ge \frac{1}{2}$ Bob maximizes this by setting $\phi = \pi/(4 k)$.
Bob prefers the resulting payoff over that of Eq.~(\ref{e-NEk}) when
\begin{equation}
\sin^2 \gamma > \frac{\$_{DD} - \$_{DC}}{\$_{DD} - \$_{CC} + \frac{1}{2} \$_{CD} - \frac{1}{2} \$_{DC}
			- \frac{1}{2} (\$_{CD} - \$_{DC})},
\end{equation}
which for the standard payoffs is never the case.
That is, Bob cannot unilaterally improve upon Eq.~(\ref{e-NEk}),
and by symmetry nor can Alice.
Thus the strategy profile $\hat{D}_{4k} \otimes \hat{D}_{4k}$
is a NE for $k \ge \frac{1}{2}$.
Following the same arguments,
$\hat{D}_{2} \otimes \hat{D}_{2}$
is a NE with
\begin{equation}
\langle \$ \rangle = \$_{CC} \, \sin^2 \gamma \, (1 - \cos 2 k \pi)/2
	\:+\: \$_{DD} \, \left[ 1 \,-\, \sin^2 \gamma \, (1 - \cos 2 k \pi)/2 \right],
\end{equation}
when $k < \frac{1}{2}$.
As $k \rightarrow 0$ or $\gamma \rightarrow 0$,
$\hat{D}_{2}$ becomes equivalent to $\hat{D}$.
The region in $(\gamma, k)$ space,
and the associated payoff,
for which $\hat{D}_2 \otimes \hat{D}_2$ is a NE
is indicated in figure~\ref{f-PDk}.
Similarly, the regions for which $\hat{C}_2 \otimes \hat{D}_2$ and $\hat{D}_2 \otimes \hat{C}_2$
are NE can be determined.
The payoffs and the region of applicability of these asymmetric NE are also indicated in figure~\ref{f-PDk}.
The defector gets the greater payoff.

We observe that by specifying the strategic space through the parameter $k$,
the presence of Nash equilibria
and, in general, the associated payoffs
can be altered.

\begin{figure}
\begin{center}
\epsfig{file=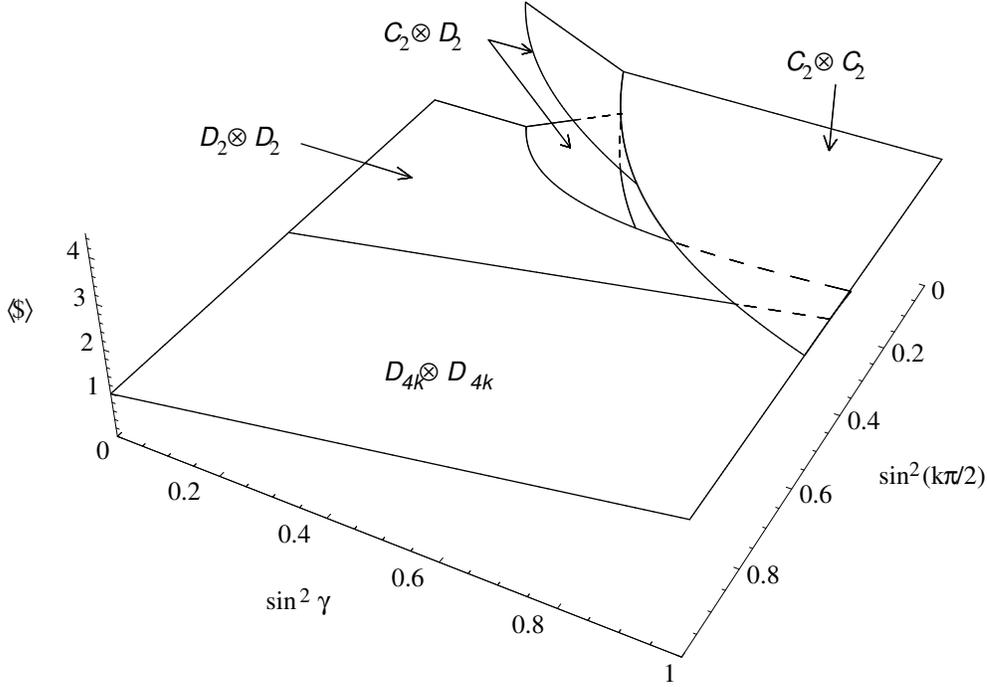, width=\textwidth}
\end{center}
\caption{The surfaces show the payoff to both players at the Nash equilibria
for a two-player quantum Prisoners' Dilemma using the strategy space $S^{(1)}_{k}$,
as a function of the entanglement and the strategy space parameter $k$.
The strategy profile $\hat{D}_{4k} \otimes \hat{D}_{4k}$ is a Nash equilibrium for $k \ge \frac{1}{2}$,
with the surface folding at $k = \frac{1}{2}$,
below which the equilibrium is $\hat{D}_2 \otimes \hat{D}_2$.
The strategy profile $\hat{C}_2 \otimes \hat{C}_2$ has a payoff of three
but is only a Nash equilibrium
for $\sin^2 \gamma \, \cos^2 \frac{k \pi}{2} \ge \frac{2}{5}$.
There is a small region where there is an asymmetric Nash equilibrium
of $\hat{C}_2 \otimes \hat{D}_2$ or $\hat{D}_2 \otimes \hat{C}_2$.
Here the defector gets a payoff of between 3 and 4,
while the cooperator must be satisfied with between 1 and 2.}
\label{f-PDk}
\end{figure}

\section{Conclusion}
We have extended the examination of two-parameter quantum strategies
in the Eisert protocol of quantum games.
In the original form presented by Eisert {\em at al.}
and subsequently taken up by a number of other authors
we have presented results for the
Nash equilibrium as a function of the entanglement parameter 
for the games of Chicken and the Battle of the Sexes,
analogous to the results previously published for Prisoners' Dilemma.
The asymmetry of the (unique) Nash equilibrium payoffs in Battle of the Sexes
already raises alarm bells about the suitability of selecting this strategic space.
By instead choosing an equivalent two-parameter strategic space where the phase factors
are arranged differently,
we show that different Nash equilibria can arise,
with different payoffs and different classical-quantum thresholds.

We generalize these results in two ways:
by examining an $N$-player quantum Prisoners' Dilemma
and by considering an extension of the two-parameter strategies
using an additional fixed parameter.
The $N$-player Prisoners' Dilemma shows a similar,
though more complex,
structure in the Nash equilibrium payoffs
to the two-player case.
For the generalized two-parameter strategies,
the position of any entanglement thresholds
between regions of differing Nash equilibria
and the associated payoffs
depend in most cases upon the (arbitrary) parameter
used in generalizing the strategy space.

These results indicate that the new equilibria are simply artifacts of the choice of strategic space,
that is, the particular slice of the space of unitary operators to which the players are restricted and,
in the absence of a physical justification for this restriction,
tell us nothing significant about the underlying game.

\section*{Acknowledgements}
Funding for APF was provided by the Australian Research Council grant number DP0559273.
LCLH is supported in part by the Australian Research Council, the Australian
government, the US National Security Agency,
the Advanced Research and Development Activity
and the US Army Research Office under contract number W911NF-04-1-0290.




\begin{thebibliography}{00}




\bibitem{eisert99} J. Eisert, M. Wilkins, M. Lewenstein,
	Phys.\ Rev.\ Lett.\ {\bf 83} (1999) 3077.
\bibitem{meyer99} D. A. Meyer,
	Phys.\ Rev.\ Lett.\ {\bf 82} (1999) 2543.
\bibitem{eisert00} J. Eisert, M. Wilkins,
	J. Mod.\ Opt.\ {\bf 47} (2000) 2543.
\bibitem{iqbal01a} A. Iqbal, A. H. Toor,
	Phys.\ Lett.\ A {\bf 280} (2001) 249.
\bibitem{du02a} J. Du, H. Li, X. Xu, M. Shi, J. Wu, R. Han,
	Phys.\ Rev.\ Lett.\ {\bf 88} (2002) 137902.
\bibitem{du02d} J. Du, X. Xu, H. Li, X. Zhou, R. Han,
      Fluct.\ Noise Lett.\ {\bf 2} (2002) R189.
\bibitem{chen03a} L. K. Chen, H. Ang, D. Kiang, L. C. Kwek, C. F. Lo,
	Phys.\ Lett.\ A {\bf 316} (2003) 317.
\bibitem{shimamura04a} J. Shimamura, S. K. {\"{O}zdemir}, F. Morikoshi, N. Imoto,
	Int.\ J.\ Quant.\ Inf.\ {\bf 2} (2004) 79.
\bibitem{ozdemir04a} S. K. \"{O}zdemir, J. Shimamura, N. Imoto,
      Phys.\ Lett.\ A {\bf 325} (2004) 104.
\bibitem{ozdemir04b} S. K. \"{O}zdemir, J. Shimamura, F. Morikoshi, N. Imoto,
      Phys.\ Lett.\ A {\bf 333} (2004) 218.
\bibitem{rapoport65} A. Rapoport, A. Chammah,
	Prisoner's Dilemma: a Study in Conflict and Cooperation,
	University of Michigan Press, Ann Arbor, 1965.
\bibitem{du01b} J. Du, X. Xu, H. Li, X. Zhou, R. Han,
      Phys.\ Lett.\ A {\bf 289} (2001) 9.
\bibitem{du03b} J. Du, H. Li, X. Xu, X. Zhou, R. Han,
      J. Phys.\ A {\bf 36} (2003) 6551.
\bibitem{benjamin01a} S. C. Benjamin, P. M. Hayden,
      Phys.\ Rev.\ Lett.\ {\bf 87} (2001) 069801.
\bibitem{flitney02c} A. P. Flitney, D. Abbott,
	Fluct.\ Noise Lett.\ {\bf 2} (2002) R175.
\bibitem{du02b} J. Du, H. Li, X. Xu, M. Shi, X. Zhou,
      Phys.\ Lett.\ A {\bf 302} (2002) 229.
\bibitem{flitney03a} A. P. Flitney, D. Abbott,
      P. R. Soc.\ London A {\bf 459} (2003) 2463.
\bibitem{marinatto00} L. Marinatto, T. Weber,
	Phys.\ Lett.\ A {\bf 272} (2000) 291.
\bibitem{benjamin00} S. C. Benjamin,
      Phys.\ Lett.\ A {\bf 277} (2000) 180.
\bibitem{nawaz04a} A. Nawaz, A. H. Toor,
      J. Phys.\ A {\bf 37} (2004) 4437.
\bibitem{schelling60} S. C. Schelling,
	The Strategy of Conflict,
	Harvard University Press, Cambridge, Massachusetts and London, 1960.
\bibitem{axelrod84} R. Axelrod, W. Hamilton,
	The Evolution of Cooperation,
	Basic Books, New York, 1984.
\end{thebibliography}
\end{document}